\documentclass[a4paper]{jpconf}
\usepackage{graphicx}
\begin{document}
\title{The Main Injector particle production Experiment (MIPP) at Fermilab}

\author{Rajendran Raja}

\address{Fermilab, P.O.Box 500, Batavia, IL 60510}

\ead{raja@fnal.gov}

\begin{abstract}
We describe the physics capabilities and status of the MIPP experiment
which is scheduled to enter its physics data taking period during
December 2004-July 2005. We show some of  the results obtained from the
engineering run that concluded in August 2004 and point out the unique
features of MIPP that make it an ideal apparatus to study
non-perturbative QCD properties.
\end{abstract}.

\section{Introduction}
The Main Injector Particle Production Experiment 
(FNAL E-907, MIPP)~\cite{mipp} is
situated in the Meson Center beamline at Fermilab. It received its
approval~\cite{proposal} in November 2001 and has installed both the experiment 
and a newly designed secondary beamline in the interim. It
received its first beams in March 2004 and concluded its engineering
run in August 2004. The physics data taking run of MIPP is scheduled
for the running period December 2004-July 2005.

MIPP is designed primarily as an experiment to measure and study in
detail the dynamics associated with non-perturbative strong
interactions. It has nearly 100\% acceptance for charged particles and
excellent momentum resolution.  Using particle identification
techniques that encompass $dE/dX$, time-of-flight~\cite{tof}, 
Multi-Cell Cerenkov~\cite{e690}
and a Ring Imaging Cerenkov (RICH)detector~\cite{rich}, 
MIPP is able to identify
charged particles at the 3$\sigma$ or better level in nearly all of
its final state phase space.

As currently envisaged, MIPP will provide events of unparalleled quality
and statistics for beam momenta ranging from 5~GeV/c to 90~GeV/c for 6
beam species ($\pi^\pm, K^\pm~ $and$~ p^\pm$.)

\section{Physics Motivation}
The primary physics motivation behind MIPP is to restart the study of
non-perturbative QCD interactions, which constitute over 99\% of the
strong interaction cross section. The available data are of poor
quality and old and are not in easily accessible form. The Time
Projection Chamber (TPC)~\cite{tpc} that is at the heart of the MIPP experiment
represents the electronic equivalent of the bubble chamber with vastly 
superior data acquisition rates. It also digitizes  the charged tracks 
in three dimensions, obviating the need for track matching across stereo views.
Coupled with the particle identification capability of MIPP, the data 
from MIPP would add significantly to our knowledge base of 
non-perturbative QCD.

One of the primary goals of the present run of MIPP is to verify a
general scaling law of inclusive particle production that states that
the ratio of a semi-inclusive cross section to an inclusive cross
section involving the same particles is a function only of the missing
mass squared ($M^2$) of the system and not of the other two Mandelstam
variables $s$ and $t$, the center of mass energy squared and the
momentum transfer squared, respectively.

Stated mathematically, the ratio

\begin{equation}
\frac{ f_{subset}(a+b \rightarrow c+X)}
{f(a+b\rightarrow c+X)}\equiv \frac{f_{subset}(M^2,s,t)}{f(M^2,s,t)}
=\beta_{subset}(M^2)
\end{equation}
{\em i.e.}, a ratio of two functions of three variables is only a
function of one of them.
This scaling relation has been shown to hold very well in a limited number of
 reactions~\cite{raja}. MIPP will test this scaling  for 36 reactions as a 
function of both $s$ and $t$ with great accuracy.

In addition to this, MIPP will acquire high quality data in liquid
hydrogen with excellent particle ID and statistics over a range of
beam momenta, which should make possible a systematic study of
exclusive reactions that is essential for testing any future theory of
non-perturbative QCD. We can also make forays into searches for exotic
resonances such as glueballs and pentaquarks. The existence of beams
of differing flavor and energies will be a great advantage in sorting
out the flavor content of any new states seen.
The other physics clients for MIPP data are nuclear and heavy ion
physics groups who are interested in  data from several nuclear
targets.  An important service measurement MIPP hopes to perform is
the measurement of particle production off the NUMI target in order to
minimize the systematics in the near/far detector ratio in the MINOS~\cite{minos}
experiment. NUMI target measurements by MIPP will also benefit the
Minerva~\cite{minerva} and the Nova~\cite{nova} experiments 
planned in the NUMI beamline in the
future. MIPP will also make measurements with proton beams off various nuclei
for the needs of proton radiography~\cite{proposal}.

Another measurement MIPP hopes to make is that of pion and proton
cross-sections off liquid nitrogen targets for the better prediction
of atmospheric neutrino fluxes. MIPP is approved for $1.3\times10^6$
spill seconds of beam to be partitioned according to the running modes
shown in table~\ref{tab1}.
\begin{table}
\caption{MIPP running scenario\label{tab1}}
\begin{center}
\begin{tabular}{c c c}
\br
Program	&Targets& Number of events ($10^6$) \\
\mr
Atm. Neutrinos	& $N_2$ &  3.00\\
NUMI & thin C, NUMI Target & 10.01\\
proton-Nucleus&	Be, Cu, Pb & 17.20\\
QCD Scaling	& $H_2$,$D_2$ & 24.40\\
proton radiography & Be, C, Cu, Pb, U & 20.79  \\
\br
\end{tabular}
\end{center}
\end{table}

\section{Experimental Setup}

We designed a secondary beam~\cite{carol} specific to our needs. The 120 GeV/c
primary protons extracted by slow resonant extraction from the
Fermilab Main Injector are transported down the Meson Center line at
Fermilab. They impinge on a 20cm long copper target producing
secondary beam particles. This target is imaged on to an adjustable momentum
selection  collimator which controls the momentum spread of the
beam. This collimator is re-imaged on to our interaction target placed
next to the TPC. The beam is tracked using three beam chambers and
 identified using two differential Cerenkovs~\cite{bckov} filled with gas, the
 composition and the pressure of which can be varied within limits 
depending onthe  beam momentum and charge.

\begin{figure}[h]
\begin{center}
\includegraphics[width=20pc]{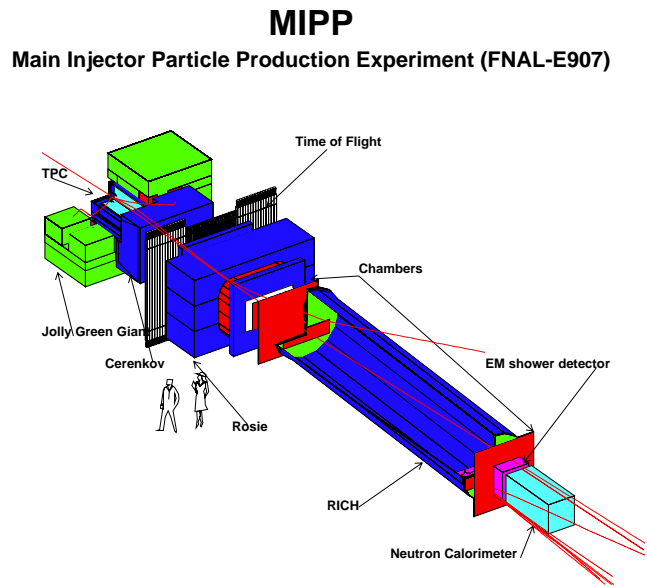}
\caption{\label{mipp}The experimental setup. The picture is a rendition in 
Geant3, 
which is used to simulate the detector}
\end{center}
\end{figure}

The study of non-perturbative QCD is of fundamental importance to our
field. However, we place too little priority on re-embarking on this
quest, largely because of the lack of a fundamental theory. MIPP hopes to
change this state of affairs by obtaining high quality data with large 
statistics, which it will publish as DST's on DVD's for public use. 
However, it was important that the cost of the detector be low in 
order for the experiment to be approved and this was achieved by refurbishing 
several existing pieces of apparatus.
Figure~\ref{mipp} shows the layout of the apparatus. The TPC sits in a
wide aperture magnet (the Jolly Green Giant) which has a peak field of
0.7 Tesla. Downstream of the TPC are a 96 mirror multi-cell Cerenkov
detector filled with $C_4F_{10}$ gas, and a time of flight system. This is
followed by a large aperture magnet (Rosie) which runs in opposite
polarity (at -0.6 Tesla)to the Jolly Green Giant to bend the 
particles back into the
Ring Imaging Cerenkov counter. The RICH has $CO_2$ as the radiator and
an array of phototubes of 32 rows and 89 columns~\cite{fire}.

Downstream of the RICH we have an electromagnetic
calorimeter~\cite{emcal} and a hadron calorimeter~\cite{hcal} to
measure forward going photons and neutrons. The electromagnetic
calorimeter will also serve as a device to measure the electron
content of our beam at lower energies, which will be useful for
measuring cross sections.

\begin{figure}[h] \begin{center} \begin{minipage}{15pc}
\includegraphics[width=15pc]{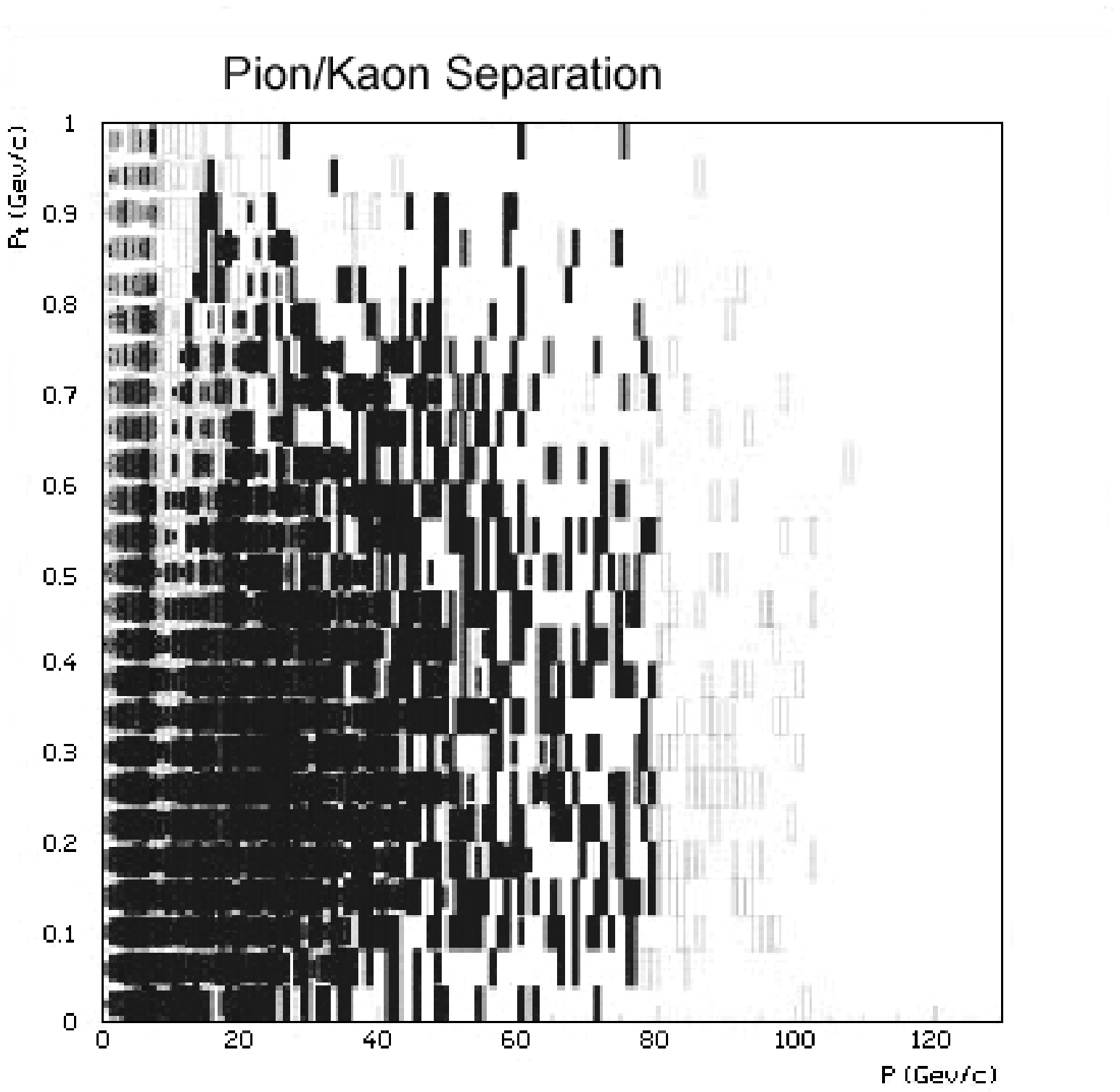} \end{minipage}
\begin{minipage}{15pc} \includegraphics[width=15pc]{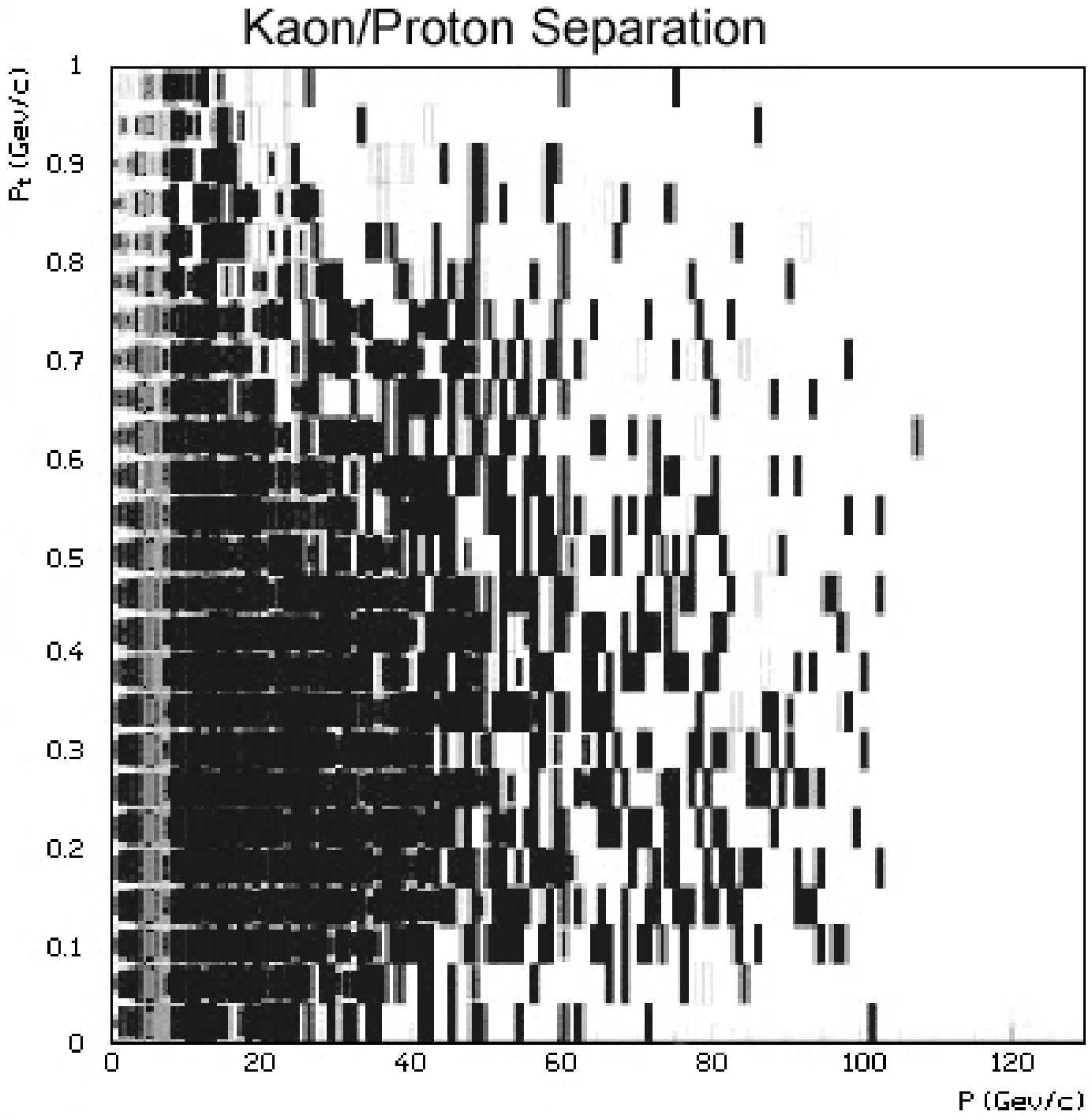}
\end{minipage} 
\caption{\label{pid}Particle ID plots for pion/kaon separation and for
kaon/proton separation as a function of the longitudinal and
transverse momentum of the outgoing final state particle. Black
indicates separation at the $3\sigma$ level or better and grey
indicates separation at the $1-3\sigma$ level.}  
\end{center}
\end{figure}

MIPP uses $dE/dx$ in the TPC to separate pions, kaons and protons for
momenta less than $\approx 1~GeV/c$, the time of flight array of
counters to do the particle id for momenta less than 2~GeV/c, the
multi-cell Cerenkov detector~\cite{e690} contributes to particle id in the
momentum range 2.5~GeV/c-7.5~GeV/c and the RICH~\cite{rich} 
for momenta higher than
this. By combining information from all counters, we get the expected
particle id separation for $K/p$ and $\pi/K$ as shown in
Figure~\ref{pid}. It can be seen that excellent separation at the
$3\sigma$ or higher level exists for both $K/p$ and $\pi/p$ over
almost all of phase space.
Tracking of the beam particles and secondary beam particles is accomplished by a set of drift chambers~\cite{chambers1} and proportional chambers~\cite{chambers2} each of which have 4 stereo layers.
\section{Engineering run results}
\begin{figure}[h]
\begin{center}
\begin{minipage}{15pc}
\includegraphics[width=15pc]{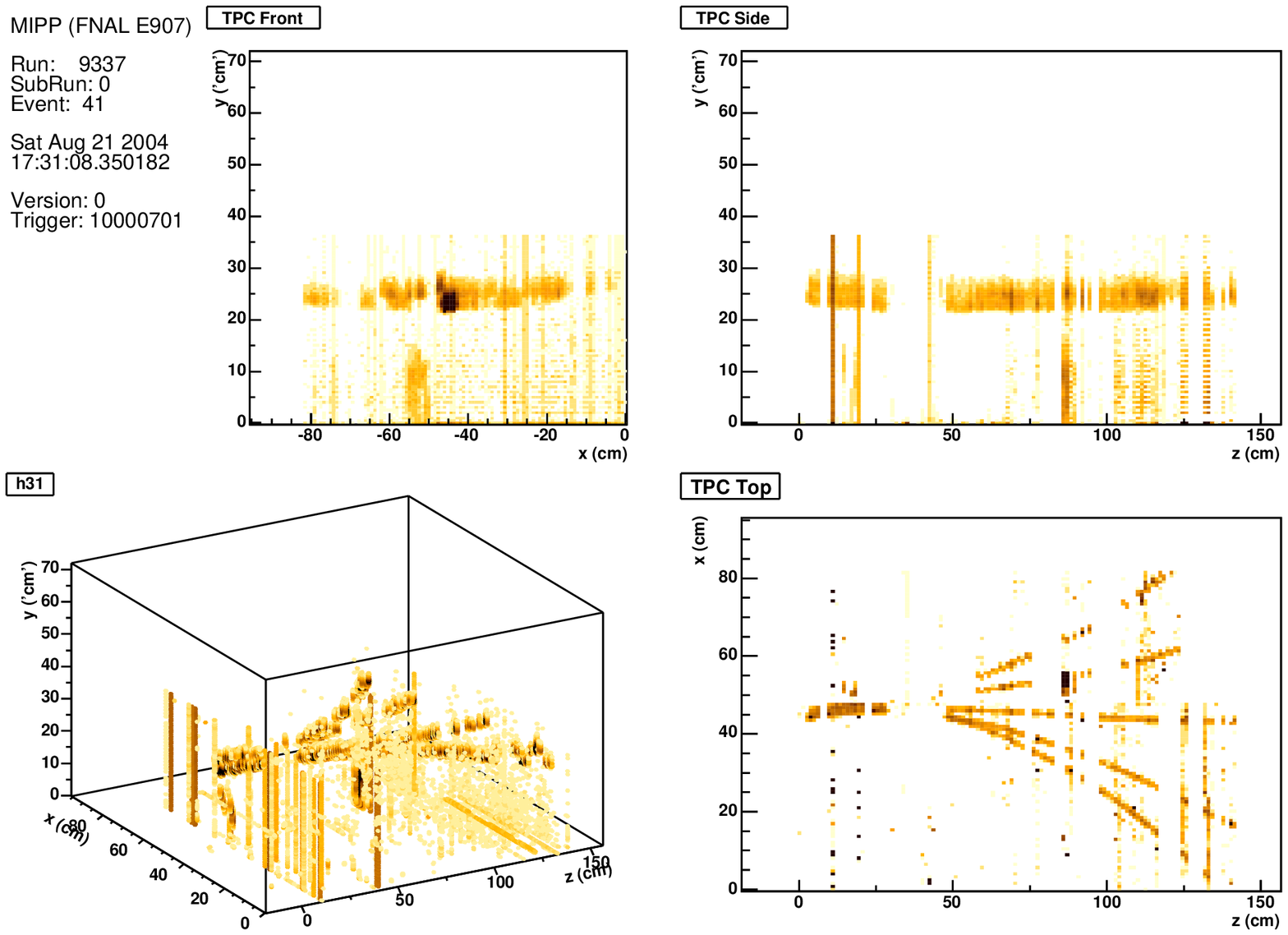}
\end{minipage}
\begin{minipage}{15pc}
\includegraphics[width=15pc]{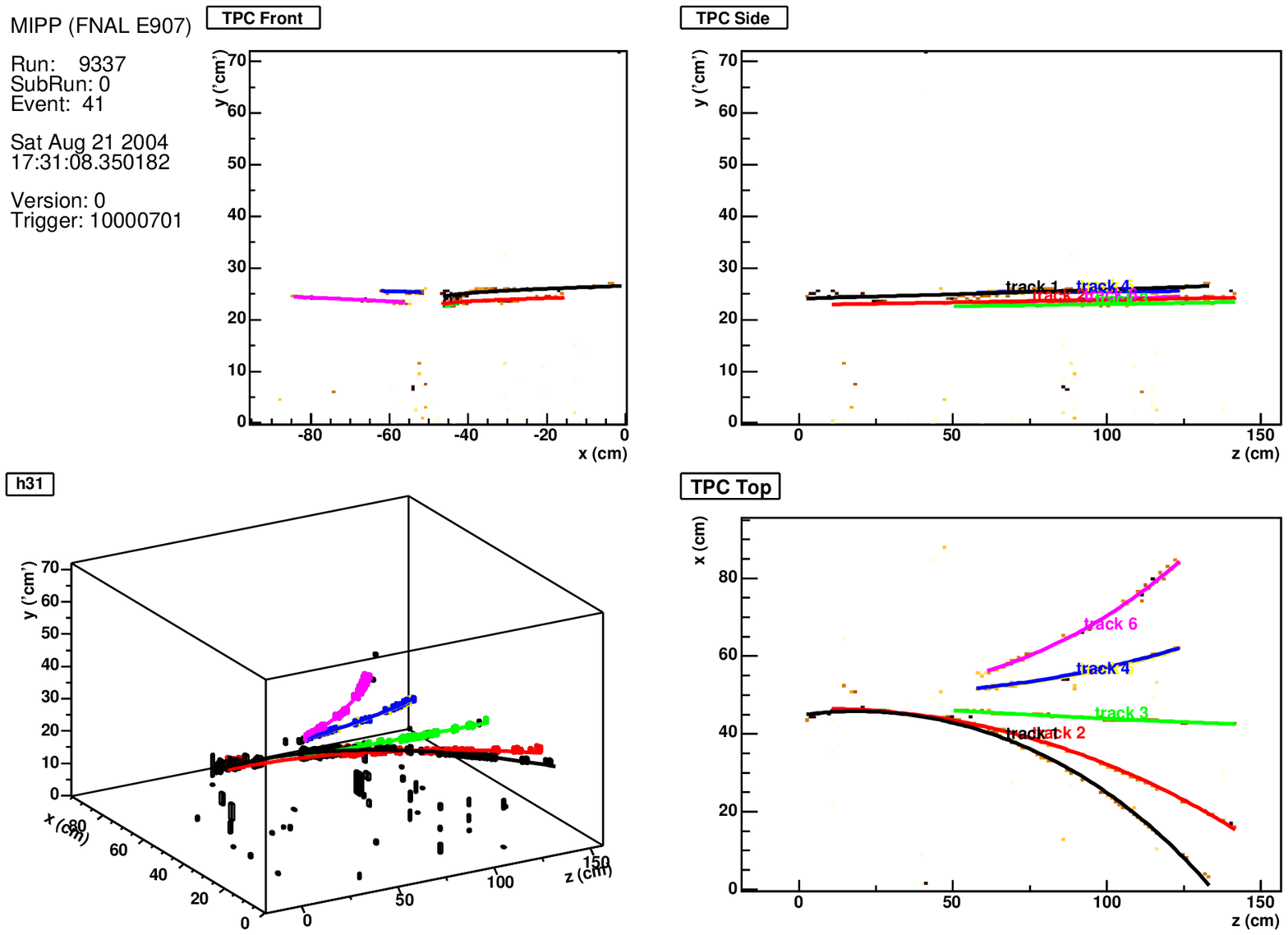}
\end{minipage}
\begin{minipage}{15pc}
\includegraphics[width=15pc]{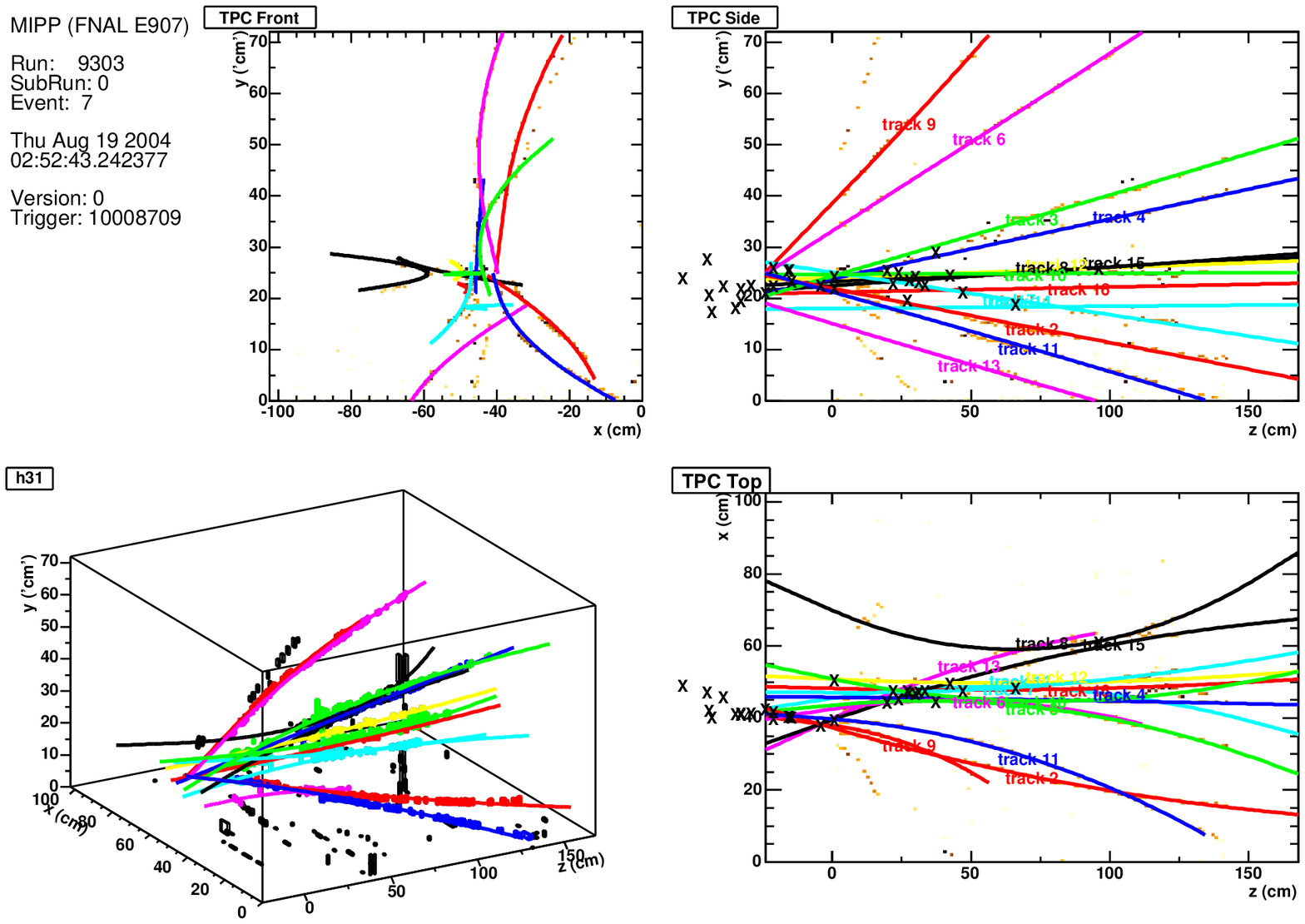}
\end{minipage}
\begin{minipage}{15pc}
\includegraphics[width=15pc]{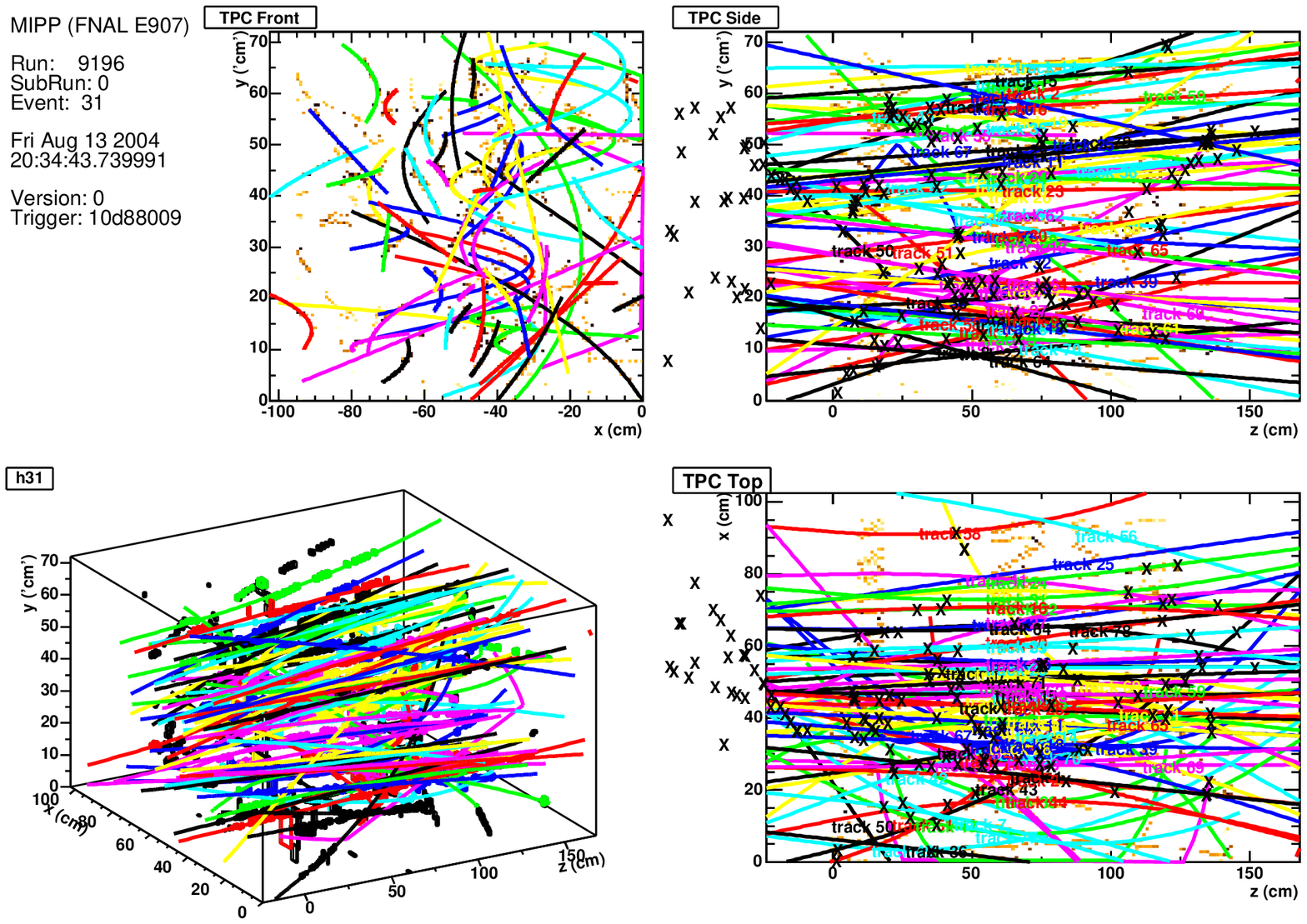}
\end{minipage}
\end{center}
\caption{\label{tpc}Pictures of TPC events. The first set of four pictures on the top left hand side show the raw digitizations of an event in 4 views and the set of four pictures on the right hand side show the tracks pattern recognized on the same event. Tracks in different views have the same color. 
The last picture is that of an upstream interaction. The offline algorithm has found all the tracks}
\end{figure}

Figure~\ref{tpc} shows the pictures of reconstructed tracks in the TPC
obtained during the engineering run. The tracks are digitized and
fitted as helices in three dimensions. Extrapolating three dimensional
tracks to the other chambers makes the pattern recognition particularly
easy.

\begin{figure}[h]
\begin{center}
\begin{minipage}{12pc}
\includegraphics[width=12pc]{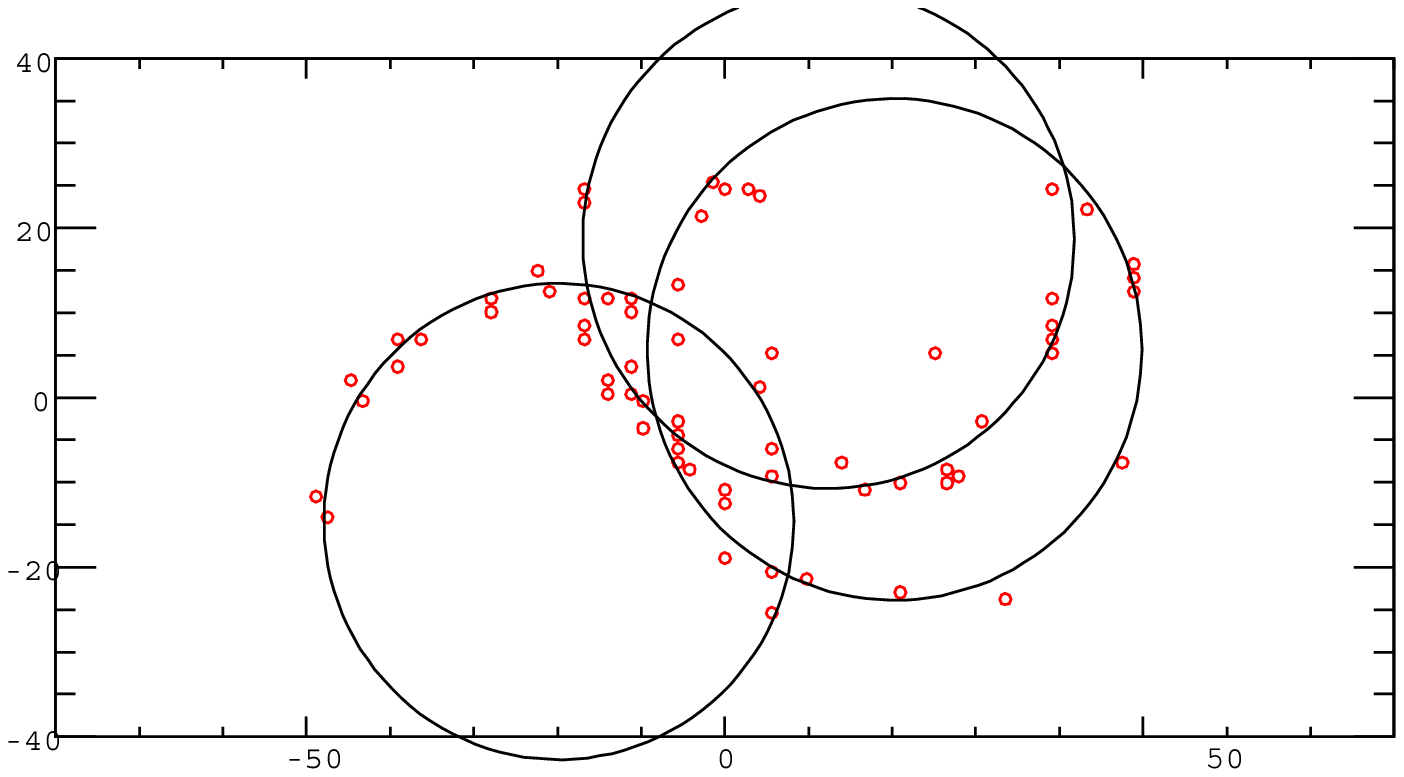}
\end{minipage}
\begin{minipage}{12pc}
\includegraphics[width=12pc]{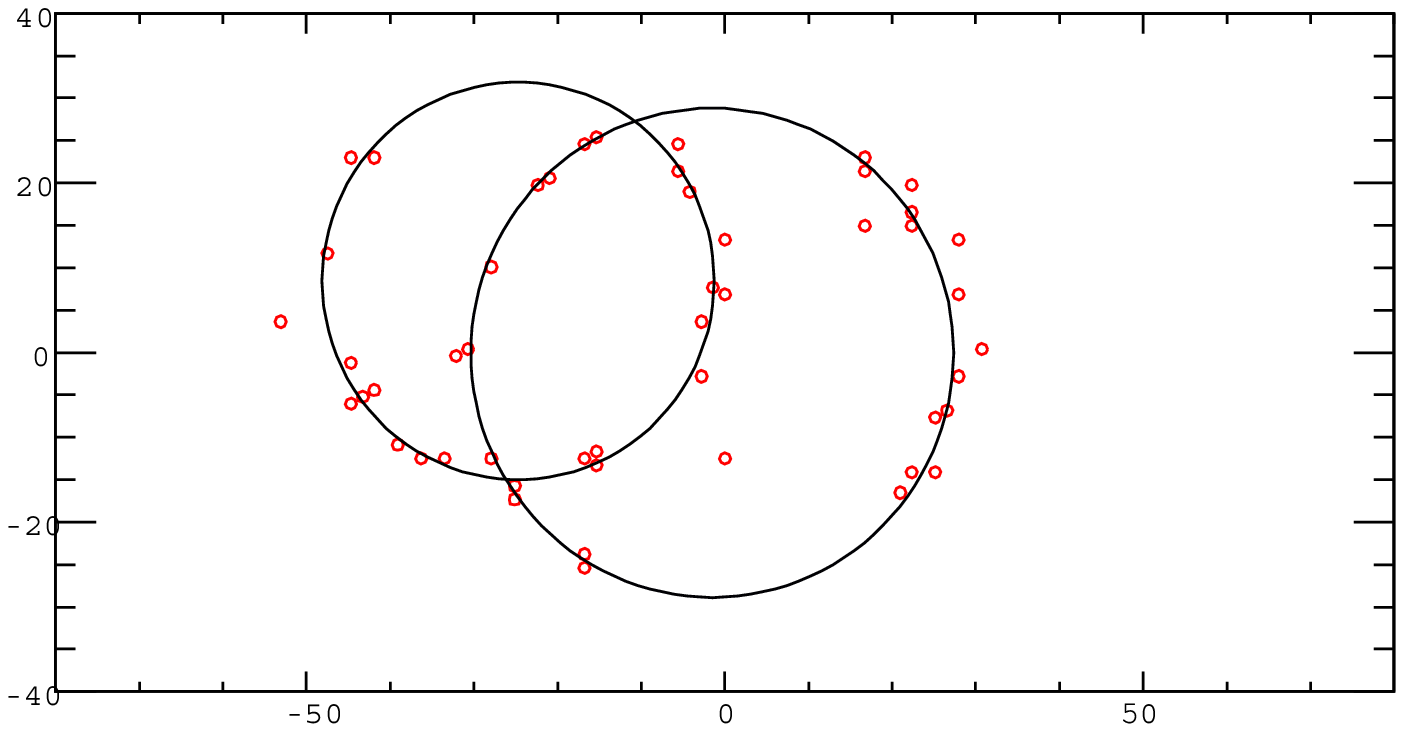}
\end{minipage}
\begin{minipage}{12pc}
\includegraphics[width=12pc]{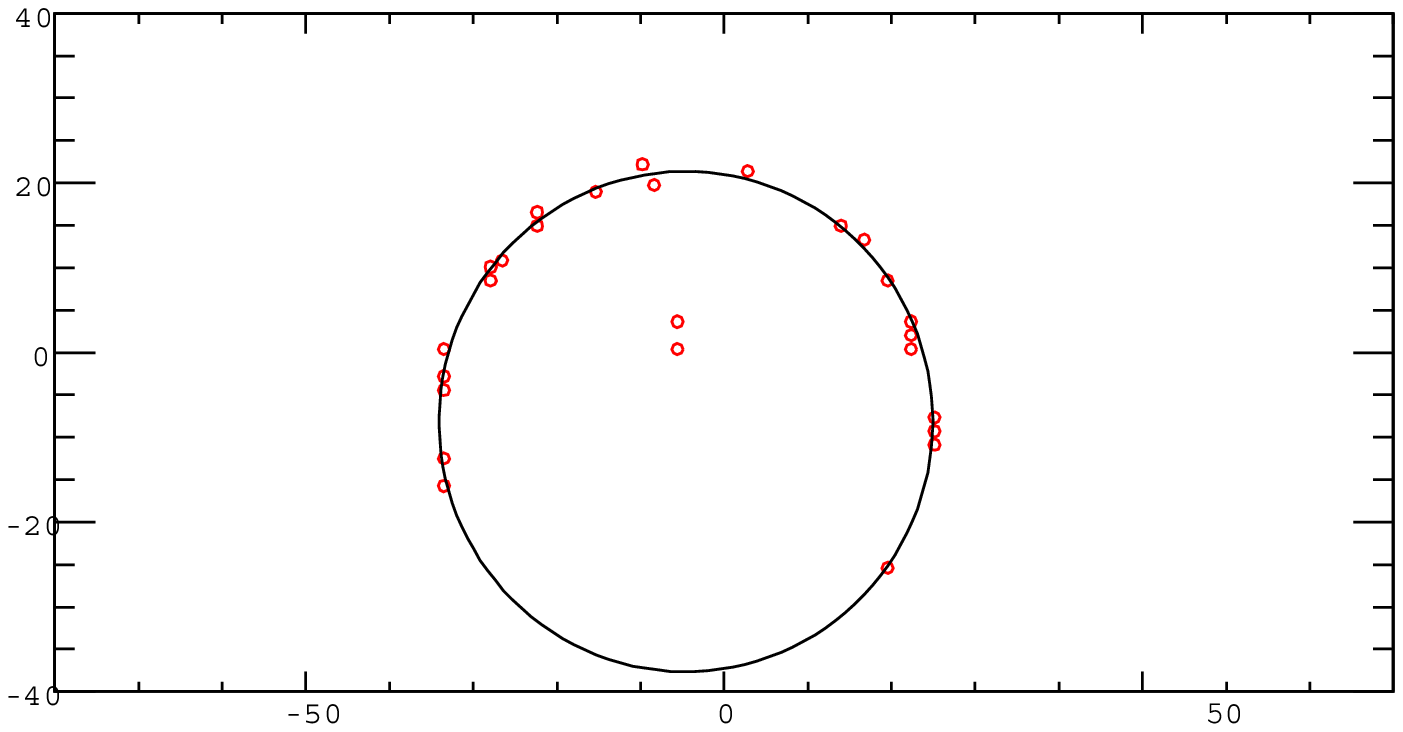}
\end{minipage}
\end{center}
\caption{\label{rings}Examples of events with rings in the RICH counter 
for a 40~GeV/c beam.}
\end{figure}

\begin{figure}[h]
\begin{center}
\begin{minipage}{15pc}
\includegraphics[width=15pc]{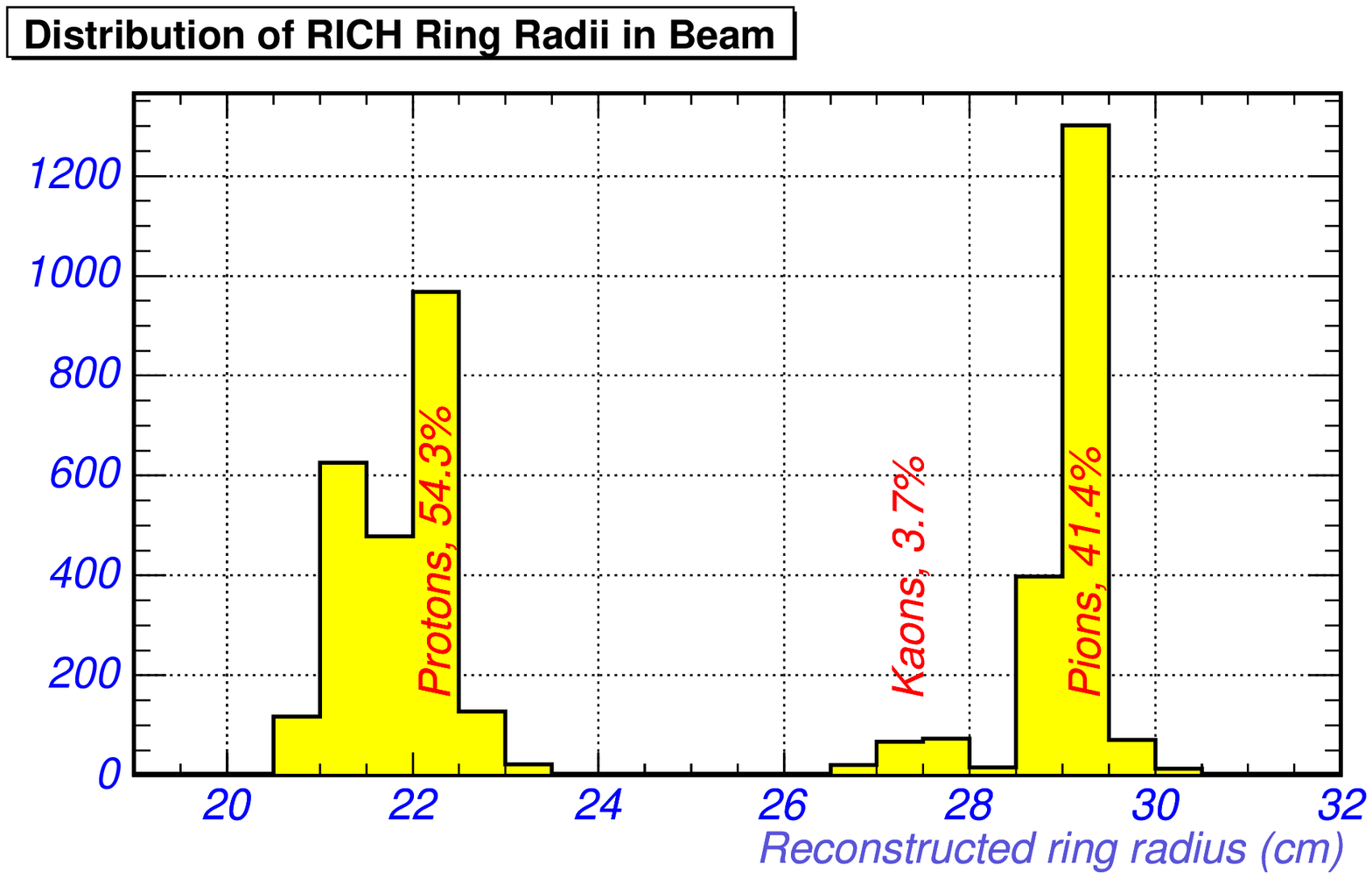}
\end{minipage}
\begin{minipage}{15pc}
\includegraphics[width=15pc]{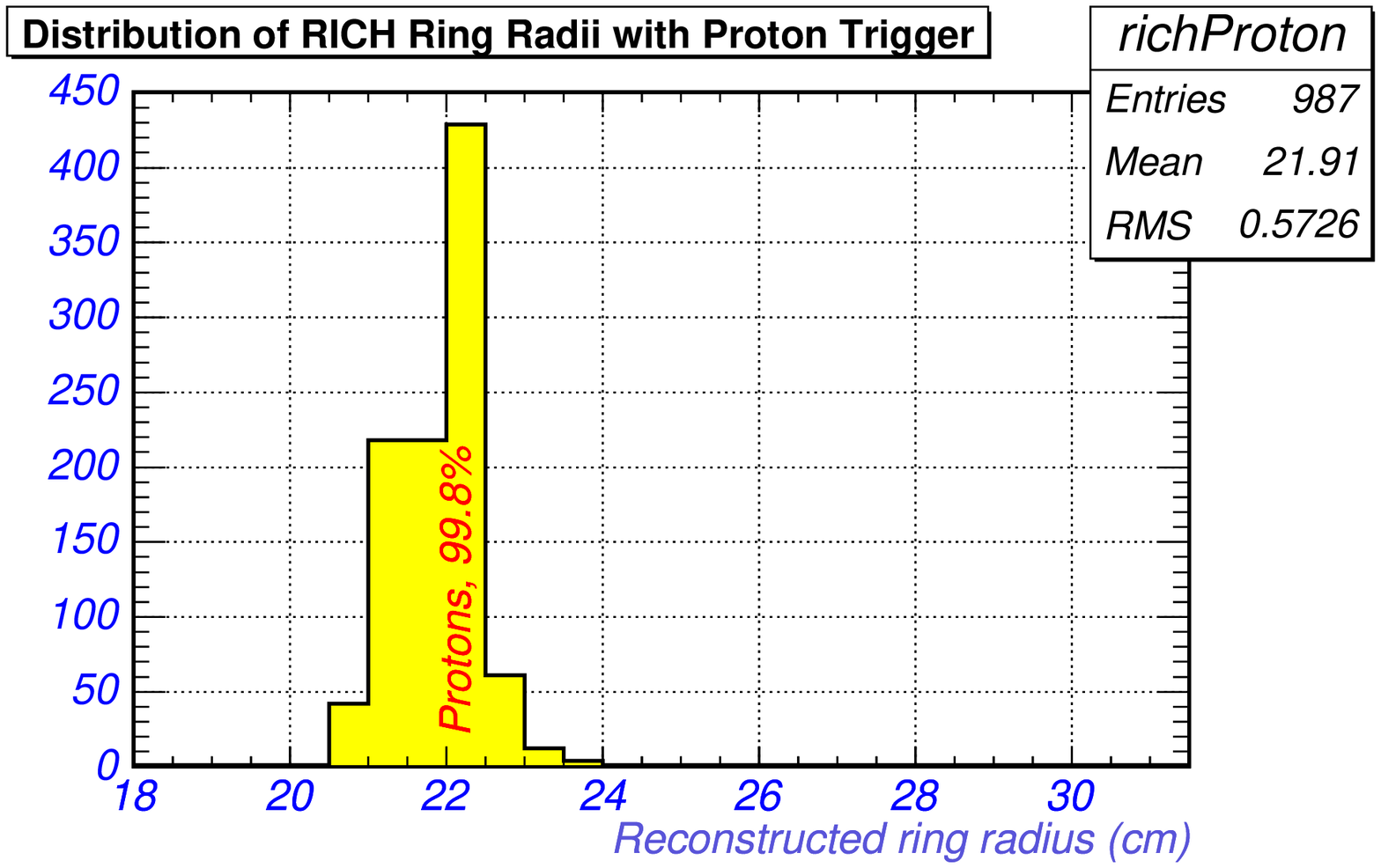}
\end{minipage}
\begin{minipage}{15pc}
\includegraphics[width=15pc]{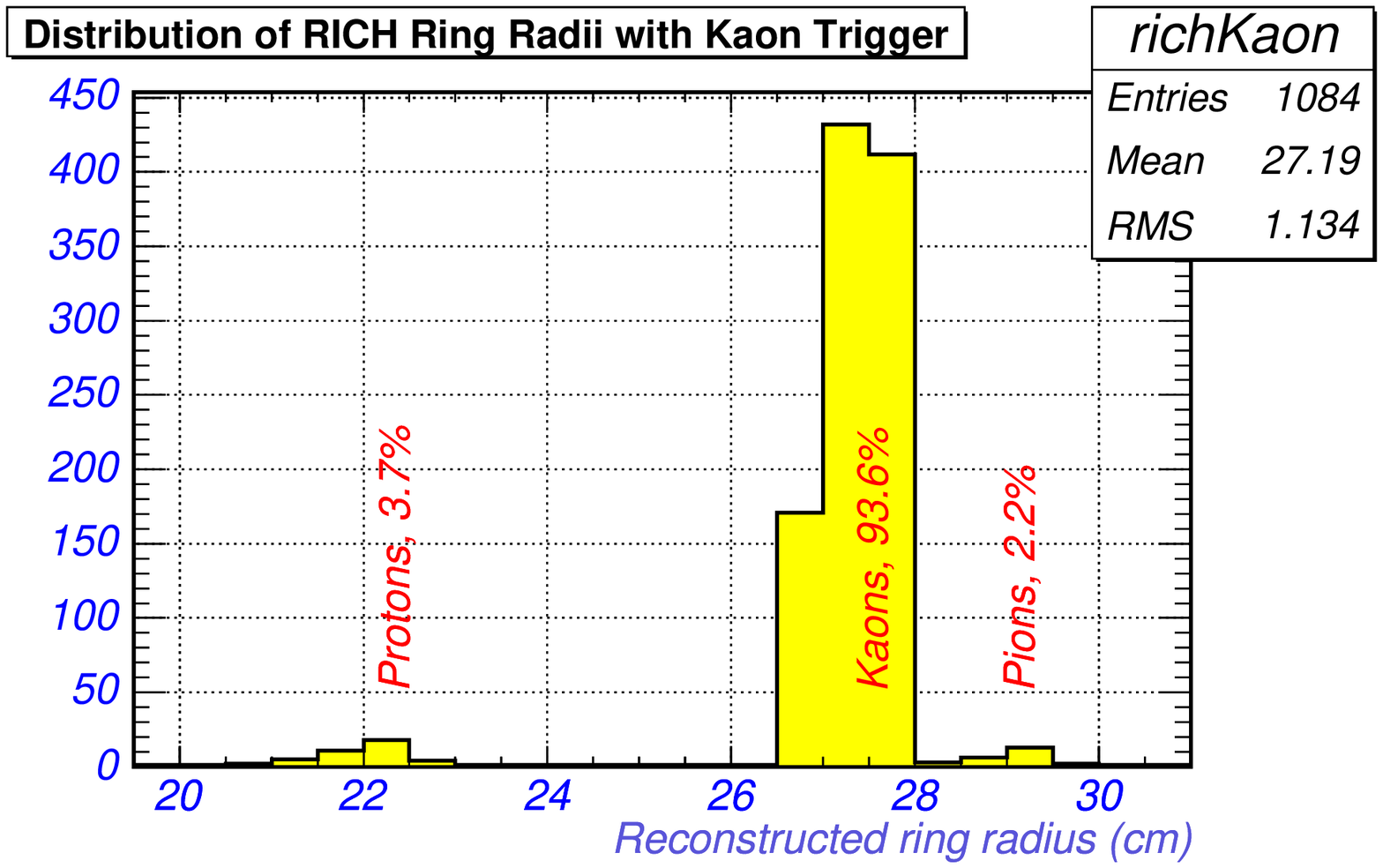}
\end{minipage}
\begin{minipage}{15pc}
\includegraphics[width=15pc]{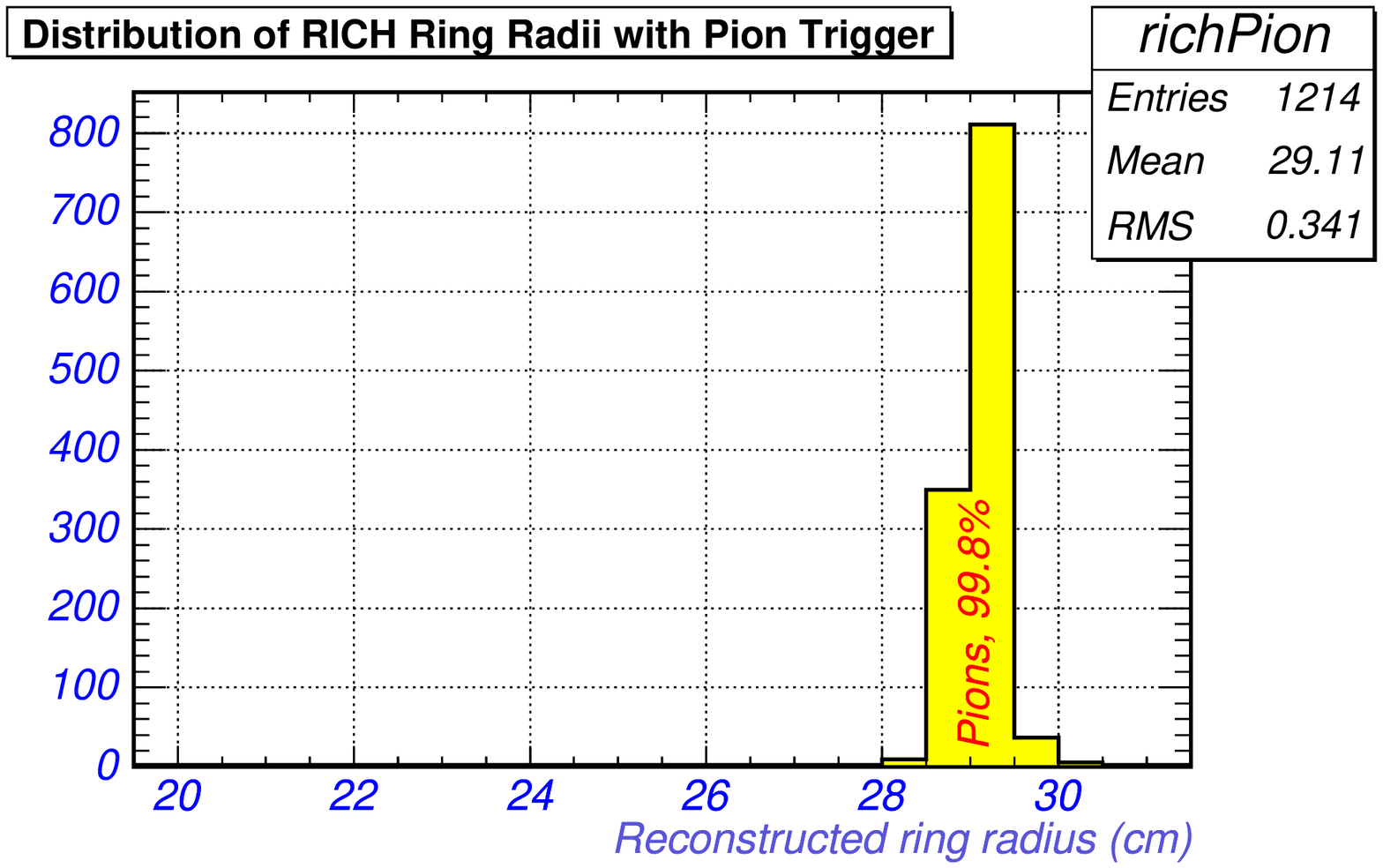}
\end{minipage}
\caption{\label{rbck}An example of a 40 GeV/c primary beam
trigger. The RICH identifies protons, kaons and pions by the ring
radii. The beam Cerenkov detectors can be used to do the same. When
the beam Cerenkov id is used, one gets a very clean separation of
pions, kaons and protons in the RICH} 
\end{center}
\end{figure}

Figure~\ref{rings} shows events with rings in the RICH counter. 
Some are due to single  beam tracks and others are due to tracks 
from interactions.
Figure~\ref{rbck} shows the histogram of ring radii for a +40GeV
secondary beam. There is clean separation between pions and kaons and
protons and their relative abundances~\cite{malensek} match expectations. Applying the
particle ID trigger from the beam Cerenkovs enables us to separate the
three particle species cleanly. The kaons which form ~4\% 
of the beam are cleanly picked out by the beam Cerenkov with very
 simple selection criteria. These can be made much more stringent 
with offline cuts to produce a very clean kaon beam. 
\section{Plans to upgrade MIPP}
At present the data taking rate in MIPP is limited by the 60HZ rate
imposed on us by the TPC electronics. These can be made considerably
faster (by at least a factor of 20)~\cite{norman} with more modern
electronics. Upgrading the TPC electronics this way accompanied by
modest upgrades to the rest of the system is expected to cost $\approx
\$0.3M$. If approved in 2005, such an upgrade can be implemented in 
time to continue the
MIPP run in 2006. This will enable MIPP to expand its present scope to
include pentaquark searches, searches for missing baryon resonances as
well as partial wave analyses using low energy beams.

\subsection{Acknowledgments}

The author wishes to thank the DoE for funding the experiment and 
Fermilab accelerator and particle divisions for continued support.


\begin{thebibliography}{9}
\bibitem{e690}
The multi-cell Cerenkov detector was initially built for Brookhaven Experiment E766 and later in Fermilab
experiment E690  and then used in several other Brookhaven experiments. In
MIPP, we fill the detector with the gas $C_4F_{10}$ which has the
appropriate refractive index at atmospheric pressure.
\bibitem{fire}
During the engineering run, we lost 20\% of the phototubes in the RICH
due to a fire in one of the bases. This does not impact
adversely on our pattern recognition, since the Cerenkov angle is large 
and there is plenty of light over most of our momentum range
\bibitem{emcal}
The electromagnetic calorimeter was fabricated by MIPP and uses lead as the 
radiator and an array of proportional tubes with 2.54 cm wire spacing 
as the readout.
It has ~10 radiation lengths and has 10 longitudinal segments.
\bibitem{hcal}
The hadron calorimeter is recycled from the HyperCP collaboration and
uses scintillator fibers embedded in iron as readout. 
It has 9.7 interaction lengths and has four 
longitudinal segmentations each of which is segmented in two transversely.
\end{thebibliography}
\end{document}